\documentclass[leqno,11pt]{article}
\usepackage{amsfonts,latexsym}
\usepackage{amsmath}
\usepackage{amscd}
\usepackage{float,amsmath,amssymb,mathrsfs,bm,multirow,graphics}
\usepackage[dvips]{graphicx}

\newcommand{\nequation}{\setcounter{equation}{0}}

\newcommand{\R}{{\Bbb R}}

\newcommand{\C}{{\Bbb C}}

\input epsf
\title{\sc A bi-Hamiltonian Supersymmetric Geodesic Equation}
\author{J. Lenells}
\date{{\small Department of Applied Mathematics and Theoretical Physics, University of Cambridge, Cambridge CB3 0WA, United Kingdom}}

\begin{document}
\maketitle

\begin{abstract} 
\noindent
A supersymmetric extension of the Hunter-Saxton equation is constructed. We present its bi-Hamiltonian structure and show that it arises geometrically as a geodesic equation on the space of superdiffeomorphisms of the circle that leave a point fixed endowed with a right-invariant metric.
\end{abstract}

\noindent
{\small{\sc AMS Subject Classification (2000)}: 37K10, 17A70.}

\noindent
{\small{\sc Keywords}: Supersymmetry, integrable systems, Hunter-Saxton equation.}

\section{Introduction} \nequation
Among all integrable systems, the KdV equation, indisputably, is the most well-studied. Ever since its solution was presented by means of the inverse scattering method \cite{GGKM}, much effort has been put into finding other physical models amenable to a similar analysis, and into developing a framework for a deeper understanding of their integrability. Geometrically, one intriguing feature of the KdV and many other integrable nonlinear evolution equations, is that they arise as Euler equations for the geodesic flow on some manifold. Since geodesic flows are not integrable in general, this raises the question of whether one may geometrically distinguish the integrable flows from the nonintegrable ones. An important class of geodesic flows, which are indeed integrable, are related to the group of diffeomorphisms of the circle, $\text{Diff}(S^1)$, or to its one-dimensional extension, the Virasoro group cf. \cite{C-K}. Indeed, the KdV equation describes the geodesic motion on the Virasoro group induced by the right-invariant metric given at the identity by the $L^2$-norm \cite{O-K}, while the $H^1$-norm on the same group gives rise \cite{M, Shkoller} to the Camassa-Holm (CH) equation \cite{C-H}---a shallow water equation which has attracted considerable interest in recent years as it exhibits more general wave phenomena than the KdV equation, such as wave-breaking and solitons with a peak at their crest cf. \cite{ACHM, C-E0, C-G-R, L3}.

One trend in the study of integrable systems has been the appearance of fermionic extensions of integrable equations. About the same time as supersymmetric methods became widely used in quantum field theory, fermionic generalizations of classical soliton equations also began to appear, two of the more well-known examples being the so-called kuperKdV \cite{Kuper} and superKdV \cite{ManinRadul, Mathieu} generalizations of the KdV equation (see \cite{ChaichianKulish, Kulish, M-P, R-K} for other examples). 
While these extensions introduce new fields in addition to the original bosonic variables, they normally also preserve most of the structure of the original equation. 
For example, it was demonstrated in \cite{O-K} that the interpretation of the KdV equation as a geodesic equation carries over to the kuperKdV system: it describes the geodesic flow on the superconformal group with respect to an $L^2$-type metric. In fact, it was shown recently \cite{D-S} that taking a general metric of $L^2$-type on the superconformal group one obtains a one-parameter family of geodesic equations which includes also the superKdV system. 

While the hope is that the interrelationship between geometry and supersymmetry would cast new light on the meaning of integrability, all the features of an integrable equation are not always preserved by a fermionic extension. 
Let us comment on the following properties of a fermionic generalization of an integrable equation: (a) it admits a bi-Hamiltonian structure (b) it is supersymmetric (c) it arises geometrically as an Euler equation for geodesic flow. 

\begin{itemize}
\item[(a)] One characteristic feature of integrable systems is the presence of two distinct, but compatible, Hamiltonian formulations. However, while this is usually considered a sufficient criterion for the integrability of a system with fermions, it appears not to be necessary. For example, only a single Hamiltonian formulation in standard form is known for the superKdV system (see \cite{O-P}). On the other hand, the kuperKdV equation has a bi-Hamiltonian formulation, but it is not supersymmetric. 

\item[(b)] The main idea of supersymmetrization, which has its origin in quantum theory, is to unify bosons and
fermions: in addition to the original commuting fields, new anti-commuting fields are introduced, the two kinds of variables being related by supersymmetry. However, fermion fields can also be incorporated independently from the boson fields, giving generalizations that are not always invariant under supersymmetric transformations. This is the case for kuperKdV.

\item[(c)] Besides the geometric formulation of the kuperKdV system mentioned above, the occurrence of fermionic extensions of the Camassa-Holm equation among the geodesic equations induced by $H^1$ type metrics on the superconformal group has also been investigated \cite{D-S}. The equations describing the geodesic motion with respect to $H^1$ type metrics define a large class of fermionic generalizations of CH. Being geodesic equations, they automatically admit at least one Hamiltonian structure---the so-called Lie-Poisson structure with Hamiltonian given by the $H^1$ metric. Moreover, within this family of $H^1$ geodesic equations, a subclass also admit a second Hamiltonian structure. However, none of the bi-Hamiltonian equations found in \cite{D-S} is also supersymmetric. Hence, the superCH system put forward in \cite{D-S} is a geodesic equation, which is supersymmetric, but for which only a single Hamiltonian formulation is known.
\end{itemize}

In addition to the Camassa-Holm equation, there recently appeared another nonlinear wave equation which also resembles the KdV equation in several respects. The Hunter-Saxton equation \cite{H-S}
\begin{equation}\label{HS}
  -u_{txx} =  2u_xu_{xx} + uu_{xxx},\quad t>0, \  x\in \R,
\end{equation}
models the propagation of orientation waves in liquid crystals, and is a bi-Hamiltonian, completely integrable system with an infinite family of conserved quantities \cite{H-Z}. Equation (\ref{HS}) also arises geometrically as a geodesic equation related to the group $\text{Diff}(S^1)$: it describes the geodesic flow on the quotient space $\text{Diff}(S^1)/S^1$ of the group $\text{Diff}(S^1)$ of orientation-preserving diffeomorphisms of the unit circle
$S^1$ modulo the subgroup of rigid rotations, endowed with the $\dot{H}^1$ right-invariant metric given by $\langle u, v \rangle = \int u_x v_x dx$ for any two vectors $u$ and $v$ in the Lie algebra $\text{Vect}(S^1)$ \cite{K-M}. 

In this note we investigate the equations describing the geodesic flow induced by a super-analogue of the $\dot{H}^1$ metric. We show that the system
\begin{equation}\label{geosystem}
\begin{cases}
-u_{txx} = 2u_xu_{xx} + uu_{xxx} + \frac{1}{2}\xi_x\xi_{xxx},
\\
-\xi_{txx} = u\xi_{xxx} + \frac{3}{2}u_x\xi_{xx} + \frac{1}{2}u_{xx}\xi_x,
\end{cases}
\end{equation}
where $u(x,t)$ and $\xi(x,t)$ are bosonic and fermionic fields, respectively, exhibits all of the above mentioned properties: it is bi-Hamiltonian and supersymmetric, and it arises geometrically as a geodesic equation on the space Super-$\text{Diff}(S^1)/S^1$ of superdiffeomorphisms of the circle that leave a point fixed.
In sections \ref{geosec}-\ref{supersec} we consider the geometric, bi-Hamiltonian, and supersymmetric structure of (\ref{geosystem}), respectively, while Section \ref{laxsec} contains a derivation of a Lax pair.
 
\section{Geodesic flow}\label{geosec}\nequation
Define the Lie algebra $\mathfrak{g}$ as the super-analogue of $\text{Vect}(S^1)$, i.e. $\mathfrak{g}$ consists of all pairs $(u(x), \varphi(x))$, where $u(x)$ and $\varphi(x)$ are bosonic and fermionic fields, respectively. Geometrically this is the superconformal algebra of contact vector fields on the $1|1$-dimensional supercircle $S^{1|1}$.
The Lie bracket is given by
\begin{equation}\label{liesuper}
\left[(u, \varphi), (v, \psi)\right] = \left(uv_x - u_x v + \frac{1}{2}\varphi \psi, u\psi_x - \frac{1}{2}u_x \psi - \varphi_x v + \frac{1}{2} \varphi v_x\right),
\end{equation}
and we introduce the $\dot{H}^1$ inner product by
\begin{equation}\label{dotH1super}
 \left\langle (u, \varphi), (v, \psi)\right\rangle = \int dx \left(u_x v_x + \varphi_x \psi\right) = \int dx\left(u A_0 v + \varphi A_1 \psi\right),
\end{equation} 
where
$$A_0 = -\partial_x^2, \qquad A_1 = - \partial_x.$$
Defining the operator $B:\mathfrak{g} \times \mathfrak{g} \to \mathfrak{g}$ by
$$\langle U, [V, W]\rangle = \langle B(U, V), W \rangle,$$
the geodesic equation can be written as $U_t = B(U,U)$ cf. \cite{A}.
Let $U = (u, \varphi)$ and $V = (v, \psi)$. Then $B(U,V) = (B_0(U,V), B_1(U,V))$, where
\begin{align}
& A_0 B_0(U,V) = -\left(2v_xA_0u + vA_0u_x + \frac{3}{2}\psi_x A_1 \varphi + \frac{1}{2}\psi A_1\varphi_x\right),
	\\
&A_1 B_1(U,V) = -\left(\frac{3}{2}v_xA_1\varphi + v A_1 \varphi_x + \frac{1}{2}\psi A_0 u\right),
\end{align}
and the geodesic equations are
$$\begin{cases}
A_0 u_t = A_0 B_0(U,V),	\\
A_1 \varphi_t = A_1 B_1(U,V).
\end{cases}$$
If we introduce $\xi$ by $\varphi = \xi_x$, these equations become (\ref{geosystem}).
Note that upon setting $\xi = 0$ in (\ref{geosystem}), the geometric formulation of equation (\ref{HS}) is recovered.

\section{Bi-Hamiltonian structure}\nequation
Introducing the variables $m = -u_{xx}$ and $\eta = - \xi_{xx}$, equation (\ref{geosystem}) admits the bi-Hamiltonian formulation
\begin{equation}\label{biham}
\begin{pmatrix} m \\ \eta \end{pmatrix}_t = J_1 \begin{pmatrix} \frac{\delta H_1}{\delta m} \\  \frac{\delta H_1}{\delta \eta} \end{pmatrix}= J_2 \begin{pmatrix} \frac{\delta H_2}{\delta m} \\  \frac{\delta H_2}{\delta \eta} \end{pmatrix},
\end{equation}
where the Hamiltonian operators are
$$J_1 = \begin{pmatrix} -\partial_x m - m \partial_x &	\frac{1}{2}\partial_x \eta + \eta \partial_x	\\
-\partial_x \eta - \frac{1}{2}\eta \partial_x 	&	-\frac{m}{2} \end{pmatrix}, \qquad J_2 =  \begin{pmatrix} \partial_x^3 &	0	\\
0	&	 \partial_x^2 \end{pmatrix}, $$
the first Hamiltonian is given by the $\dot{H}_1$ inner product according to
$$H_1 = \frac{1}{2}\langle U, U \rangle = \frac{1}{2} \int dx \left(u_x^2 +  \xi_{xx}\xi_x\right),$$
and the second Hamiltonian is
$$H_2 =\frac{1}{2} \int dx \left(uu_x^2 - u\xi_x \xi_{xx}\right).$$
Indeed, for a functional $F$, the variational derivatives $\frac{\delta F}{\delta m}$ and $\frac{\delta F}{\delta \eta}$ are defined by
$$\frac{d}{d\epsilon} \biggr|_{\epsilon = 0} F[m + \epsilon \delta m, \eta + \epsilon \delta \eta] = \int dx \left(\frac{\delta F}{\delta m} \delta m + \frac{\delta F}{\delta \eta} \delta \eta\right).$$
It follows from the definitions of $m$ and $\eta$ that
$$A_0 \begin{pmatrix} \frac{\delta F}{\delta m} \\ \frac{\delta F}{\delta \eta} \end{pmatrix}
= \begin{pmatrix} \frac{\delta F}{\delta u} \\  \frac{\delta F}{\delta \xi} \end{pmatrix}.$$
Hence, the formulation (\ref{biham}) is easily verified using that
$$\begin{pmatrix} \frac{\delta H_1}{\delta u} \\  \frac{\delta H_1}{\delta \xi} \end{pmatrix}
= A_0\begin{pmatrix} u \\ \xi_x \end{pmatrix}, \qquad
\begin{pmatrix} \frac{\delta H_2}{\delta u} \\  \frac{\delta H_2}{\delta \xi} \end{pmatrix}
= -\frac{1}{2} \begin{pmatrix}  u_x^2 + 2uu_{xx} + \xi_x \xi_{xx}	\\   
2u\xi_{xxx} + 3u_x \xi_{xx} + u_{xx} \xi_x
\end{pmatrix}.$$
Notice that when $\xi = 0$, the bi-Hamiltonian structure reduces to that of the Hunter-Saxton equation cf. \cite{H-Z}.
Let us also point out that equation (\ref{geosystem}) has a Lagrangian formulation corresponding to the Hamiltonian $H_2$. It is the Euler-Lagrange equation for the action
$$S = \int \int dt dx \left( u_tu_x - \xi_t \xi_{xx} + uu_x^2 - u\xi_x \xi_{xx}\right).$$

\section{Supersymmetry}\label{supersec}\nequation
A simple computation shows that equation (\ref{geosystem}) is invariant under the supersymmetry transformation
\begin{equation}\label{susytransformation}  
  \delta u = \tau \xi_x, \qquad \delta \xi = \tau u,
\end{equation}
where $\tau $ is an odd parameter. The supersymmetry can also be established by formulating the calculations in Section \ref{geosec} in terms of superfields. We define a superderivative $D$ by $D = \partial_\theta + \theta \partial_x$, where $\theta$ is an odd coordinate. Letting $U = u + \theta \varphi$ and $V = v + \theta \psi$, the Lie bracket (\ref{liesuper}) can be written as
$$\left[U, V\right] = UD^2 V - VD^2 U + \frac{1}{2} (DU)(DV),$$
and the $\dot{H}^1$ inner product becomes
$$\langle U, V \rangle = \int dxd\theta (D^2U)(DV).$$
The superspace bilinear operator $\hat{B}(U, V)$ satisfies
$$-D^3\hat{B}(U, V) = V (D^5 U) + \frac{1}{2} (DV) (D^4 U) + \frac{3}{2} (D^2V )(D^3 U).$$
Introducing a fermionic superfield $M(x, \theta)$ by $M = - D^3 U = -\varphi_x + \theta m$, the geodesic flow equation $U_t = \hat{B}(U, U)$ is
\begin{equation}\label{superspacesystem}
  M_t = U(D^5 U) + \frac{1}{2} (DU) (D^4 U) + \frac{3}{2} (D^2 U )(D^3 U).
\end{equation}
This equation is by construction invariant under the supersymmetry transformation (\ref{susytransformation}). The component equations of (\ref{superspacesystem}) give back system (\ref{geosystem}).

\section{Lax pair}\label{laxsec}\nequation
A long but straightforward computation shows that equation (\ref{superspacesystem}) is the condition of compatibility of the linear system
\begin{equation}\label{lax}
\begin{cases} 
  D^3 G = \frac{1}{2\lambda} M G,
  	\\
  G_t = \frac{1}{2}U_x G - \frac{1}{2}(DU)(DG) + (\lambda - U)G_x.
\end{cases}
\end{equation}
where the even superfield $G$ serves as an eigenfunction and $\lambda \in \C$ is a spectral parameter.

Let us briefly indicate how the Lax pair (\ref{lax}) was derived. We first consider how the existence of a recursion operator leads to a Lax pair in the purely bosonic case. By definition, a recursion operator $R$ for an evolution equation $m_t = K[m]$ satisfies 
\begin{equation}\label{recursionlax}  
  R_t = [K', R],
\end{equation}
where $K'[m]$ is the Frech\'et derivative of the operator $K$ cf. \cite{Fsymmetries}. Equation (\ref{recursionlax}) is the compatibility condition of
\begin{equation}\label{varphilax} 
 \begin{cases} 
 R \varphi = \lambda \varphi,
	\\
 \varphi_t = K' \cdot \varphi,
 \end{cases}
 \end{equation}
The system (\ref{varphilax}) provides a Lax pair for the evolution equation in terms of the `squared eigenfunction' $\varphi$ cf. \cite{F-A}. We expect the existence of a simpler Lax pair expressed in terms of an appropriate `square root' of $\varphi$. 

This procedure can be applied to the Hunter-Saxton equation (\ref{HS}) as follows. Since equation (\ref{HS}) admits the recursion operator $R = (m\partial_x + \partial_x m) \partial_x^{-3}$, where $m = -u_{xx}$ (see \cite{H-Z}), we obtain an $x$-part of the form
$$(m\partial_x + \partial_x m) \partial_x^{-3}\varphi = \lambda \varphi.$$
Moreover, letting $\varphi = \partial_x^3(\psi^2)$, this equation is implied by the $x$-part of the standard Lax pair for (\ref{HS}) given by
\begin{equation}\label{psilax}
\begin{cases}
  \psi_{xx} = \frac{1}{2\lambda} m \psi,
		\\
  \psi_t = \frac{1}{2}u_x \psi + (\lambda - u)\psi_x.
  \end{cases}
\end{equation} 

We now show how (\ref{lax}) can be derived by following analogous steps in the presence of supersymmetry.\footnote{An alternative approach would be to start with the bosonic Lax pair (\ref{psilax}) and simply search for the most general fermionic extension compatible with supersymmetry. The current method is more systematic and provides additional guidance.}
 In terms of $M$, the bi-Hamiltonian formulation (\ref{biham}) reads
$$M_t = K_1 \frac{\delta H_1}{\delta M} = K_2 \frac{\delta H_2}{\delta M},$$
where
\begin{align*}
 H_1 = \frac{1}{2}\int dx d\theta (U M),
\qquad 	 H_2 = \frac{1}{2}\int dx d\theta (D^2U)(D U) U,
\end{align*}
and
$$K_1 = -\frac{1}{2} \left[M \partial_x + 2 \partial_x M + (D M)D\right], \qquad  K_2 = D^{5}.$$
We consider the recursion operator
$$R = -K_1 K_2^{-1} =  \frac{1}{2} \left[M \partial_x + 2 \partial_x M + (D M)D\right]D^{-5}.$$
In analogy to the above bosonic case, we expect
\begin{equation}\label{RD5F}  
  R F = \lambda F,
\end{equation}
to be the $x$-part for a Lax pair of (\ref{superspacesystem}). The even superfield $F$ serves as a `squared eigenfunction'. Letting $F = D^5(G^2)$, we find that the equation
\begin{equation}\label{D3Gguess}  
  D^3 G = \frac{1}{2\lambda} M G,
\end{equation}
implies (\ref{RD5F}). This suggests that equation (\ref{D3Gguess}) is an appropriate $x$-part. In components, with $G= g + \theta \nu$ and $M = -\varphi_x + \theta m$, (\ref{D3Gguess}) reads
\begin{equation}\label{xpartcomponents}
\begin{cases}
  g_{xx} = \frac{1}{2\lambda} (m g + \varphi_x \nu)
  	\\
  \nu_x = - \frac{1}{2\lambda} \varphi_x g,
  \end{cases}
\end{equation}
which agrees with the $x$-part of (\ref{psilax}) when the fermionic fields are set to zero. 

In order to find the corresponding $t$-part, we seek a Lax pair of the form
\begin{equation}\label{seeklax}
\begin{cases}
  D^3 G = \frac{1}{2\lambda} M G  	\\
  G_t = AG + BDG + CG_x,
\end{cases}
\end{equation}
where $A,C$ are even superfields and $B$ is an odd superfield to be determined. The compatibility condition of (\ref{seeklax}) is
$$\left(\frac{1}{2\lambda} M G\right)_t = D^3\left(AG + BDG + CG_x\right).$$
We expand the derivatives, use (\ref{seeklax}) to replace all occurences of $G_t$ and $D^3G$, and identify coefficients of $G, DG,$ and $G_x$, to arrive at the three equations
\begin{align}\label{eq1}
  M_t =&2\lambda (DA_x) + (DB) M + (DC)(DM) + C_x M - B(DM) + C M_x,
 	\\\label{eq2}
 0 = &DB_x - \frac{1}{2\lambda}(DC)M + A_x + \frac{1}{\lambda}B M,
 	\\\label{eq3}
 0 =& DC_x + DA - B_x.
\end{align}
Letting
$$A = \frac{1}{2}U_x, \qquad B = - \frac{1}{2}D U, \qquad C =  \lambda - U,$$
equations (\ref{eq2}) and (\ref{eq3}) are identically satisfied, while equation (\ref{eq1}) is equivalent to (\ref{superspacesystem}).

\section{Concluding remarks}\nequation
We have constructed a supersymmetric, bi-Hamiltonian extension of equation (\ref{HS}) and shown that it arises geometrically as the geodesic equation with respect to the $\dot{H}^1$ metric (\ref{dotH1super}) on Super-$\text{Diff}(S^1)/S^1$. Later, the bi-Hamiltonian structure has been used to derive a Lax pair. The equation sets itself apart from the super-geodesic equations previously obtained using the $L^2$ and $H^1$ metrics in that it is both supersymmetric and bi-Hamiltonian: the kuperKdV system which is induced by the $L^2$ metric \cite{O-K} is not supersymmetric \cite{D-M}, while the superCH system put forward in \cite{D-S} is supersymmetric but appears not to be bi-Hamiltonian. 

Finally, let us point out that the geometric interpretation of equation (\ref{HS}) is particularly simple in that the underlying space $\text{Diff}(S^1)/S^1$ is isometric to an open subset of an infinite-dimensional $L^2$-sphere, so that the geodesics are simply segments of the big circles on the sphere \cite{Lsphere}. It would be interesting to investigate whether this point of view has an extension to the supersymmetric system (\ref{geosystem}).

\bigskip
\noindent
{\bf Acknowledgement} {\it The author thanks Professor O. Lechtenfeld for helpful discussions. This work was carried out while the author was supported by a Marie Curie Intra-European Fellowship.}

\bibliography{is}

\end{document}